\documentclass[submission,copyright,creativecommons]{eptcs}
\providecommand{\event}{AREA 2022} 
\usepackage{breakurl}             
\usepackage[inline]{enumitem}
\usepackage{tikz}
\usepackage{float}
\usepackage{xspace}
\usepackage{subcaption}
\usepackage{listings}
\newcommand{\vesna}{{VEsNA}\xspace}

\title{Towards VEsNA, a Framework for Managing Virtual Environments via Natural Language Agents}
\author{Andrea Gatti and Viviana Mascardi
\institute{Department of Computer Science, Bioengineering, Robotics and Systems Engineering (DIBRIS) \\ University of Genova, Italy}
\email{s4496922@studenti.unige.it, viviana.mascardi@unige.it}
}
\def\titlerunning{Towards VEsNA, a Framework for Managing Virtual Environments via Natural Language Agents}
\def\authorrunning{A. Gatti \& V. Mascardi}

\begin{document}
\maketitle

\begin{abstract}
Automating a factory where robots are involved is neither trivial nor cheap. Engineering the factory automation process in such a way that return of interest is maximized and risk for workers and equipment is minimized, is hence of paramount importance. Simulation can be a game changer in this scenario but requires advanced programming skills that domain experts and industrial designers might not have.

In this paper we present the preliminary design and implementation of a general-purpose framework for creating and exploiting Virtual Environments via Natural language Agents (\vesna). \vesna
takes advantage of agent-based technologies and natural language processing to enhance the design of virtual environments.  
The natural language input provided to \vesna is understood by a chatbot and passed to a cognitive intelligent agent that implements the logic behind displacing objects in the virtual environment. In the \vesna vision, the intelligent agent will be able to reason on this displacement and on its compliance to legal and normative constraints. It will also be able to implement what-if analysis and case-based reasoning. Objects populating the virtual environment will include active objects and will populate a dynamic simulation whose outcomes will be interpreted by the cognitive agent; explanations and suggestions will be passed back to the user by the chatbot.
\end{abstract}

\section{Introduction and Motivation}
\label{introMot}


Factory automation is a safety-critical task that can help significantly in increasing production, but that does not come free of charge. Robots may be very expensive and their impact over the production pipeline is not always predictable. For this reason,  many factory automation commercial software applications exist, including Computer Aided Design (CAD) tools and simulators\footnote{See, for example, \url{www.fastsuite.com/solutions-products/market-specific/factory-automation},  \url{store.indusuite.com/products/software/}, \url{www.createasoft.com/automation-simulation}, \url{https://new.siemens.com/global/en/products/automation/topic-areas/simulation-for-automation.html}.}.
Those applications are highly customized for factory automation. They are complete and efficient in that domain but require advanced programming skills to be used, that might prevent some categories of users from taking advantage of them -- for example, small factories where industrial designers do not have sufficient computer science background, or where the price of those commercial software applications cannot be afforded. In those situations a cheap and ``extremely-easy-to-setup'' virtual twin of the factory along with the robots and the workers therein might bring many benefits, even if less effective and complete than the simulations generated by ad-hoc toolkits.

Consider, for example, the following scenario.
%
%
\begin{figure}[ht]
    \centering
	\includegraphics[scale=.4]{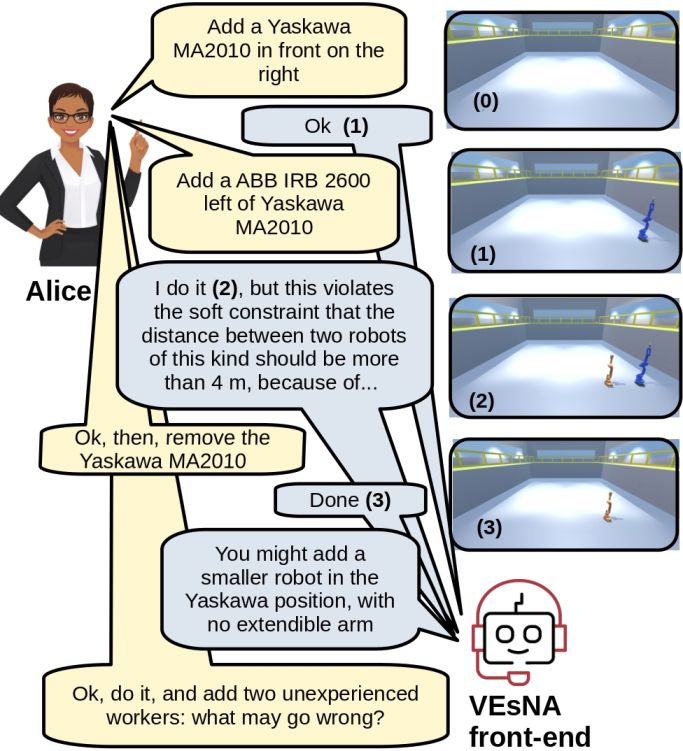}
	\caption{Example of visionary interaction between a user and \vesna.}
	\label{fig:motivation}
\end{figure}
Alice runs a small factory where various types of metallic components for industrial automation are assembled.  She is a visionary businesswoman and in her vision safety deserves the first place.
Alice got a funding for making her factory more efficient and productive by installing new robots. She wants to identify the safest configuration of robots in the industrial building that she rented but her limited budget does not allow her to pay for the license of a commercial CAD or simulation tool.
Alice is aware of specific issues raised by the robots she is going to buy and install, that may generate hard and soft constraints to meet, and of general legal requirements that may prevent her from making some choices. She also knows that robots may undergo malfunctioning, but that also human workers may not behave in the correct way. To ensure safety she would like to anticipate -- or at least to understand -- what might go wrong with different configurations of the robots. For this reason, she needs a (possibly unexpensive and open) tool that allows her to:
\begin{itemize}
\item[1] displace elements (different kinds of robots, furniture, ...) into a virtual environment via a user-friendly interface based on natural language interaction, in order to identify the best configuration of the building;
\item[2] check that displaced objects meet the hard and soft safety and legal constraints related to their position and interactions;
\item[3] provide a natural language explanation of why some constraints are not met by a given configuration, and suggest alternatives.
\end{itemize}
These three goals might be achieved by adding a natural language interface on top of any existing Computer Aided Design tool, but Alice needs something more sophisticated. Indeed, she would like to insert dynamic elements into the virtual environment, so that the result is not just a static rendering, but a running simulation. In particular, she wants to add virtual humans, and she wants to anticipate what may happen in case of unexpected or wrong maneuvers made by the human workers. The tool that Alice is looking for, should also allow her to
\begin{itemize}
\item[4] run simulations in the virtual environment, get statistics about them, and explain her -- using a natural language interface -- what may go wrong.
\end{itemize}
But this is still not enough. Once the best configuration will be devised, Alice wants to train her employees to move and act in the factory, before they start working in the real environment.  The last feature of the tool Alice needs, is to
\begin{itemize}
\item[5] allow workers to {\em enter} the virtual and dynamic environment, interact with robots therein, learn what is safe and what is not, and get the most effective and realistic training with natural language explanations.
\end{itemize}

\noindent The tool Alice is looking for should
\begin{itemize}
\item[1] understand command issued in natural language; those commands might range from the simplest ``Add a Yaskawa MA2010 in front on the right'' to the more sophisticated ``Add workers with some profile (experienced, unexperienced, reliable, etc...)''; the tool should provide a chatbot-like natural language interface;
\item[2] be aware of rules (normative, physical, domain-dependent) and be able to verify whether a given configuration -- that may include robots and workers -- meets them; the tool should be able to reason on facts represented in a symbolic way (``a robot is placed in this position and an unexperienced worker is placed in that position''), on their logical consequences, and on the rules that may be broken by them;
\item[3]  be able to synthesise natural language explanations of its logical reasoning flow, besides reasoning and understanding natural language;
\item[4] ensure a one-to-one correspondence between facts representing symbolic knowledge a\-me\-na\-ble for logical reasoning and objects placed in a virtual environment -- equipped with a realistic graphical interface -- where dynamic behaviors may be added: the tool should be suitable for running realistic simulations; 
\item[5] allow the virtual environment to be made available as a virtual reality for training purposes, with no extra effort.  
\end{itemize}

To the best of our knowledge no such a tool exists: this paper moves the initial steps towards it by describing \vesna, an implemented general-purpose framework for managing Virtual Environments via Natural language Agents\footnote{\vesna is freely available to the community from \url{https://github.com/driacats/VEsNA}.}. At its current stage of development, \vesna is far from offering the services that Alice needs. Nevertheless, the three technologies it builds upon have the {\em potential} to cope with all her needs, and some promising results have already been achieved in the natural language interaction, reasoning, explanation, simulation, training in the virtual reality dimension, although no integration among these capabilities has been provided yet. In fact, \vesna exploits
	\begin{enumerate*}
		\item[(i)] \textbf{DialogFlow}\footnote{So far, DialogFlow is the only non open-source technology in \vesna ; we are considering to substitute it with an open-source equivalent application, Rasa, to make \vesna completely open source.} for building a chatbot-like interface,
		\item[(ii)] \textbf{JaCaMo} for integrating knowledge coming from the interaction with the user into a cognitive, rule-driven agent able to reason about this knowledge and to provide human-readable explanations;
		\item[(iii)] \textbf{Unity} for building the dynamic virtual environment, and letting human users immerse in it.
	\end{enumerate*}

The goal that \vesna aims at achieving in the future is to provide Alice and many other users with an integrated environment for managing virtual environments via natural language, cognitive agents able to support decision making thanks to their reasoning and explanatory capabilities: an example of interaction between Alice and \vesna is depicted in Figure \ref{fig:motivation}.

The problem that \vesna solves in its current state is to integrate DialogFlow, JaCaMo and Unity into a single framework and to provide the means to {\em displace the elements relevant for finding the best configuration of the industrial building into  a simulated environment via a user-friendly interface based on natural language interaction}.

The paper is organized in the following way. Section~\ref{sec:background} introduces the three technologies  \vesna builds upon and motivates our confidence that filling the gap between the \vesna -as-is and the \vesna -to-be is technically feasible. Section~\ref{sec:related} positions our contribution w.r.t. the state-of-the-art. Section~\ref{sec:intonate} illustrates the envisioned \vesna interaction and workflow and presents an example in the factory automation domain. Section~\ref{sec:conclusions-future-work} concludes the paper and discusses future developments.

\section{Background}
\label{sec:background}

		\paragraph{DialogFlow.}
	    DialogFlow \cite{dialogflow} is a \textit{lifelike conversational AI with state-of-the-art virtual agents} developed by Google.
		It allows users to create personal chatbots, namely \textbf{conversational agents} equipped with \textbf{intents}, \textbf{entities}, and \textbf{fulfillment}.

		During a conversation between humans, a human speaker can utter different types of sentences, each one with a different intentional meaning.
		That meaning can be identified as the {\em intent} of that sentence.
		For example, if someone says \textit{"Hello!"}, the hearer can infer that the (friendly, socially-oriented) intent is of greeting her. If someone says \textit{"Go away!"}, the (unfriendly, command-like, action-oriented) intent is of having some concrete action performed by the hearer -- namely, moving away. The speech act theory \cite{austin1962how,searle_1979} provides a theoretical and philosophical basis for this intention-driven communication model.
		In order to explain the map between sentences and intents to the chatbot agent, the agent's developer should provide examples of sentences that convey that intent, for each intent that is relevant for the application.

		Real sentences are much more complex than \textit{"Hello!"} and \textit{"Go away!"}: they usually add some contextual information to the main intent. For example, in the \textit{"My name is Bob"} sentence, the speaker's (friendly, socially-oriented) intent is to introduce himself,  with the additional information about his name. The agent must be instructed that the  \textit{name} is something it should remember about persons, when they introduce themselves.
		This goal can be achieved by creating an {\em entity} \textit{Person} with a field \textit{name}, that we refer to as a ``parameter''.
		In the set of sentences associated with the intent, the string that identifies the name should be highlighted by the agent's developer, to let the agent learn how to retrieve the person's name inside sentences tagged with the ``introduce himself'' intent.

		Once the sentence intent has been understood and the parameters have been retrieved, the agent must provide a meaningful and appropriate answer. By default, the answer is a fixed sentence whose only variable parts are those related with parameters identified in the input sentence.
		For example, if the user types \textit{"Hello! My name is Bob"}, the agent can identify the name and answer something like \textit{"Hi Bob! Nice to meet you!"}, but nothing more advanced.
		This is where {\em fulfillment} comes into play.
		The fulfillment is a sort of help from home: if the agent cannot answer messages for some specific intent, those messages are forwarded to an external, specialized source that is waiting.
		Fulfillment provides a field where the user can insert the \textsc{url} address of the service to query.
		The service at that address will be consulted only for those intents that require it; in that case, DialogFlow will wait for the answer and will forward it to the user.

	\paragraph{JaCaMo.}
	   JaCaMo \cite{boissier2013multi,boissier2020multi}
	   \textit{"is a framework for Multi-Agent Programming that combines three separate technologies, each of them being well-known on its own and developed for a number of years so they are fairly robust and fully-fledged"} \cite{jacamowebsite}.

		The three technologies that JaCaMo integrates  are:
		\begin{enumerate*}
			\item[(1)] \textbf{Jason} \cite{bordini2007programming}, for programming autonomous agents characterized by mentalistic notions like beliefs, goals, desires, intentions, and able to reason;
			\item[(2)] \textbf{CArtAgO} \cite{ricci2006cartago,ricci2011environment}, for programming environment artifacts;
			\item[(3)] \textbf{MOISE} \cite{hubner2007developing}, for programming multi-agent organisations.
		\end{enumerate*}
		In our work, we will use only Jason and (in an indirect way) CArtAgO.
		By using JaCaMo we can build a multiagent system along with its environment.
		JaCaMo agents follow the Beliefs-Desires-Intentions (BDI) model \cite{rao1995bdi,georgeff1998belief} and are implemented in Jason, that is a variant of the logic-based AgentSpeak(L) language \cite{rao1996agentspeak}.
		The Jason elements that are more relevant for programming one individual, cognitive agent are:
		\begin{itemize}
			\item \textbf{Beliefs:} the set of facts the agent knows,
			\item \textbf{Goals:} the set of goals the agent wants to achieve,
			\item \textbf{Plans:} the set of pre-compiled, operational plans the agent can use to achieve its goals.
		\end{itemize}
		To make an example, a simple reactive agent that turns on the light either when someone enters the room, or when someone issues the "turn light on" voice command, or any other command with the same semantics, would need to know
		\begin{enumerate*}
			\item[(1)] either when someone enters the room or when the "turn light on" command has been issued, and
			\item[(2)] how to turn on the light.
		\end{enumerate*}
		The first piece of knowledge is a \textit{belief} (something the agent knows about the world either because it is informed of that fact by some other agent via communication, or because that fact about the environment is sensed via sensors or provided by artifacts, or because the agent itself generates that piece of knowledge) and the second one is a \textit{plan} (a recipe to achieve some goal via a sequence of  operational steps).
		Considering the first way to have light on, triggered from people entering the room, the scenario might involve a sensor that checks the presence of people in the room and triggers the addition or removal of belief from the agent's belief base depending on its measurement.
		Last, the agent has two \textit{goals}:
		\begin{enumerate}
			\item turn on the light when the right condition is met,
			\item turn off the light when no-one is there.
		\end{enumerate}
		We assume that the agent's plans offer instructions on how to turn on the light, and how to turn it off. So, depending on its beliefs about the presence of people in the room (that in turn depend on the sensor's outcome, or on a command been issued), the agent decides to adopt one plan or the other, in order to achieve its goal.


		The agent needs a way to communicate with the environment.
		Usually, CArtAgO artifacts are used for this purpose.
		In our example, we assume to have an artifact controlling the sensor and another one receiving information from a Dialogflow agent that listens for the user's commands.
		The bridge between DialogFlow and JaCaMo is built with \textbf{Dial4Jaca}~\cite{Dial4Jaca,DBLP:conf/paams/EngelmannDKBCPB21} that provides a set of CArtAgO artifacts that run throughout the execution of the JaCaMo agent.
		Dial4Jaca starts a listener and has the knowledge to receive and interpret messages from DialogFlow.
		These messages are not the ones the user writes on the chat, but a standardized and structured format that DialogFlow uses when it has to use fullfillments.
		Such messages contain the intent name and the entities' parameters identified (e.g., the object the user is talking about in the scene).
		When the messages are received by a JaCaMo agent, they are parsed and then added to the agent's beliefs, using Jason's syntax.
		Finally, the addition of such new belief inside the JaCaMo agent's belief base will cause a reaction driven by the most suitable plan among those in the agent's plan base.
		The use of Jason and JaCaMo paves the way to supporting those advanced features mentioned in Section \ref{introMot} namely sophisticated reasoning capabilities and goal-driven planning \cite{DBLP:books/daglib/0001819,DBLP:conf/atal/RicciBHC18,DBLP:journals/tcs/GabrielPBAB20}, exploitation of formal and semi-formal methods to implement monitoring and safety checks \cite{DBLP:conf/atal/BordiniFPW03,DBLP:conf/dalt/AnconaDM12,DBLP:conf/atal/FerrandoDA0M18,area2022}, explainability, also in connection with DialogFlow thanks to Dial4JaCa \cite{DBLP:conf/bracis/EngelmannCPB21,PAAMS2022}.


		\paragraph{Unity.}
		Unity \cite{unity} is a cross-platform game engine that allows developers to create scenes, and add objects to such scenes by dragging and dropping them from a palette to the scene.
		Objects inserted into a scene can be more or less realistic, and may enjoy or not physical properties, depending on what is needed in the application domain of choice.
		When an object is put inside a scene, a script written in C\# can be attached to it.
		From that script, it is possible to instantiate other objects inside the scene, and modify (resp., destroy) those already placed there.
		Unity may allow users to have a controllable running simulation where elements have some degree of autonomy \cite{DBLP:conf/paams/BiagettiFM20}, and to turn that simulated environment into a virtual reality \cite{DBLP:conf/vr/JeraldGWHCK14}.

\section{Related Work}
\label{sec:related}

The idea of integrating BDI agents into game engines like Unity is extremely recent and almost unexplored; surprisingly, the more general idea that existing game engines and simulation platforms are suited to act as platforms for building MASs - especially when learning capabilities are required - is still also poorly explored.

One of the first works dealing with the general vision dates back to 2014 \cite{BECKERASANO2014452}
and presents a multi-agent system based on Unity 4 that allows simulating three-dimensional way-finding behavior of several hundreds of airport passengers on an average gaming personal computer. Although very preliminary, that work inspired successive research where -- however -- the focus was not on the use of a game engine as a general-purpose MAS platform, but rather on specific problems that a 3D realistic simulation raises such as signage visibility to improve pathfinding \cite{MOTAMEDI2017248}, and on ad-hoc simulations \cite{articleJiacheng,YILDIZ2020106597,doi:10.1080/14498596.2020.1787253}.

In~\cite{DBLP:journals/corr/abs-1809-02627} Juliani et al. move a step forward generalization and present a taxonomy of existing simulation platforms that enable the development of learning environments that are rich in visual, physical, task, and social complexity. In their paper, the authors argue that modern game engines are uniquely suited to act as general platforms and examine Unity and the open source Unity ML-Agents  Toolkit\footnote{\url{https://github.com/Unity-Technologies/ml-agents}} as case studies.

When moving to the integration of BDI-style agents into game engines, one of the first works to mention is the PRESTO project \cite{busetta2016applying}, spanning 2013-2016, whose outcomes were significantly constrained by the immaturity of game engines like Unity, besides the immaturity of platforms for BDI agents.

In the Master Thesis by N. Poli dating back 2018 \cite{poli2018game}, simple BDI agents were implemented using a lightweight Prolog engine, tuProlog \cite{DBLP:journals/scp/DentiOR05}, that overcame some limitations of UnityProlog\footnote{\url{https://github.com/ianhorswill/UnityProlog}.}, an existing Prolog intrpreter compatible with Unity3D. A roadmap to exploit game engines to model {MAS} that also discusses the results achieved in  \cite{poli2018game} has been published by S. Mariani and A. Omicini in 2016 \cite{DBLP:conf/woa/MarianiO16a}.

Similarly to the work by N. Poli, the work by Mar\'\i n-Lora et al. \cite{MARINLORA2020102732} describes a game engine to create games as multi-agent systems where the behaviour specification system is based on first-order logic and is hence closer to a declarative approach than a purely procedural programming language.

Simulations that exploit Unity as the engine and visualization tool and the BDI model as a reference for implementing individual agents have been developed in a few domains including large urban areas \cite{8584878}, fire evacuation \cite{paschal2022developing}, first aid emergency \cite{10.1177/1046878119865913}, gas and oil industry \cite{matoso2020agent}, bushfires in Australia \cite{wai2021autonomous}. Those works are driven by an application to simulate and lack re-usability in other contexts.

The work closest to ours, at least in its final goal of creating a general-purpose extension of Unity that integrates BDI agents, is hence the one by A. Brännström and J. C. Nieves in~\cite{emas2022}. There, the authors introduce UnityIIS, a lightweight framework for implementing intelligent interactive systems that integrate symbolic knowledge bases for reasoning, planning, and rational decision-making in interactions with humans. This is done by integrating Web Ontology Language (OWL)-based reasoning \cite{owl:2004} and Answer Set Programming (ASP)-based planning software \cite{DBLP:books/sp/99/Lifschitz99,DBLP:books/sp/Lifschitz19} into Unity.
Using the components of the UnityIIS framework, the authors developed an Augmented
Reality chatbot following a BDI model: beliefs are the agent’s internal knowledge of its environment,
which is updated during the interaction by getting new observations; desires are
goals that the agent aims to fulfill, which are updated during the interaction by
reflecting upon new beliefs; intentions are what goals the agent has chosen to
achieve, selected in a deliberation process and used for generating a plan.
The belief of the agent is represented in an OWL ontology, as also suggested by other authors in the past \cite{DBLP:conf/dalt/MoreiraVBH05,DBLP:journals/wias/MascardiABBR14}. The UnityIIS framework enables belief revisions (OWL/ASP file updates) at run time by interweaving the agent’s control loop with the Unity game loop. The Unity game loop does cyclic update iterations on a
given frequency. During each frame, external inputs are processed, game status
is updated, and graphics are redrawn.

Albeit sharing some similarities, there are also differences between \vesna and UnityIIS. The main one is that UnityIIS does not integrate JaCaMo with Unity: the BDI model is used as a reference, but is not implemented using a standard agent programming language as Jason. Rather, UnityIIS relies on OWL and ASP as languages for modelling knowledge and declarative behaviour of cognitive agents. The second one is that the chatbot described  in~\cite{emas2022} is an example of application of UnityIIS, in the same way as factory automation is an example of application of \vesna. In \vesna, the chatbot is one of the three pillars of the framework, and not just an application. What we might borrow from the UnityIIS model, is the adoption of OWL to model knowledge and reason about it in an interoperable,  portable way. What UnityIIS might borrow from \vesna, is the closer integration of a standard framework for BDI agents into the system, JaCaMo, along with all its libraries and add-ons, rather than the adoption of more generic logic-based languages like ASP and OWL.


\section{Envisioned \vesna Interaction and Workflow}
\label{sec:intonate}
    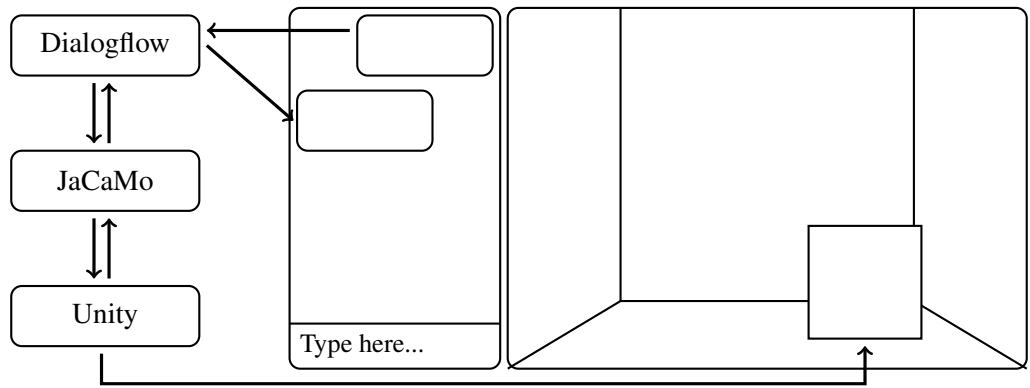
\begin{figure}[ht]
		\centering
		\begin{tikzpicture}[thick]

			\draw[rounded corners] (-4.9, 2.1) rectangle ++ (2.8, 4.8);
			\draw (-4.9, 2.7) -- (-2.1, 2.7);
			\draw (-4.9, 2.4) node[right]{\small Type here...};
			\draw[rounded corners] (-4, 6) rectangle ++ (1.8, 0.8);
			\draw[rounded corners] (-4.8, 5) rectangle ++ (1.8, 0.8);

			\draw[rounded corners] (-2, 2.1) rectangle ++ (6.9, 4.8);
			\draw (-2, 2.1) -- (-.5, 3) -- (3.4, 3) -- (4.9, 2.1);
			\draw (-.5, 3) -- (-.5, 6.9);
			\draw (3.4, 3) -- (3.4, 6.9);

			\draw[->, very thick] (-4.1, 6.6) -- (-6, 6.6);
			\draw[->, very thick] (-6, 6.4) -- (-4.85, 5.4);
			\draw[rounded corners] (-8.6, 6) rectangle node{Dialogflow} ++ (2.5, 0.8);
			\draw[->, very thick] (-7.5, 5.9) -- (-7.5, 5.1);
			\draw[<-, very thick] (-7.3, 5.9) -- (-7.3, 5.1);
			\draw[rounded corners] (-8.6, 4.2) rectangle node{JaCaMo} ++ (2.5, 0.8);
			\draw[->, very thick] (-7.5, 4.1) -- (-7.5, 3.3);
			\draw[<-, very thick] (-7.3, 4.1) -- (-7.3, 3.3);
			\draw[rounded corners] (-8.6, 2.4) rectangle node{Unity} ++ (2.5, 0.8);
			\draw[->, very thick] (-7.4, 2.3) -- (-7.4, 1.9) -- (2.75, 1.9) -- (2.75, 2.4);
			\draw[fill=white] (2, 2.5) rectangle ++ (1.5, 1.5);
		\end{tikzpicture}
		\caption{\vesna architectural scheme.}
		\label{scheme}
	\end{figure}

In this section, we describe how a user may interact with \vesna. Figure~\ref{scheme} reports a high-level architectural scheme of \vesna.
On the \vesna screen, the user can find a chat and a pre-built scene made with Unity representing whatever he/she needs, based on the application domain; the scene has at least one empty floor.
The user can describe how the scene is organized via the chat, using natural language\footnote{So far we have run experiments with English only, but given that DialogFlow natively supports multiple languages, \url{https://cloud.google.com/dialogflow/es/docs/agents-multilingual}, letting the user talk in his/her own language should be also possible.} and so telling the \vesna agent what to do.
For example, the user could say\footnote{So far, interactions are via a textual interface only; nonetheless, to plug a voice-to-text component into the \vesna control flow should not raise any technical issues.} \textit{``Add that object in the scene!''}.
This would make the object appear inside the scene. However, the translation from the message to the graphics animation requires various intermediate steps.

First, the DialogFlow agent identifies the name of the object the user wants to add along with its position in the scene. This information is automatically extracted from the user's sentence through natural language processing.
In more detail, the position can be described by the user in two different ways. The first is by using a \textbf{global positioning} on the entire space available.
In the current implementation, the space is subdivided into 9 available positions. Hence, the user can say something like \textit{``Ok agent, put that object in front on the right''} and the agent will understand what it is expected to do.
While the global positioning may work well with small scenarios, it is likely to be too coarse grained with big ones; however, this may represent the ``first'' positioning of objects inside the scene. Indeed, other objects may be added later on by referring to relative positions (w.r.t. already present objects in the scene). The user can add new objects that \textbf{reference} those already in scene.
In that case, it would be possible to say \textit{``Ok, I told you that there is that object on right right. Well, behind it there is this other object''}.
This is performed by giving a unique name to each object added inside the scene. Such name can be used any time it is needed to identify an object as reference.
Note that, when a user adds an object, the answer contains the \textit{reference name} for it.

Furthermore, while Dialogflow can identify every object name inserted by the user, an error will be generated if it is not one of the available items.
To make the system work, the user has to insert the name of the object in Unity, in such a way that DialogFlow can recognize it.
For the moment, this applies also to identifiers of already added objects: they must be those provided by Unity and communicated to the user.
We are working to overcome this limitation by introducing scopes that will allow the user a more natural description of the scene, making prior knowledge available and supporting the ability to refer the last added objects without having to remember their  identifiers.

Once the user's sentence has been processed by DialogFlow, the resulting intent is generated. In the case of the addition of a new object inside the scene, the intent \textit{AddObject} is used and trained to identify \textit{who} is the object to add and \textit{where} to add it.
It has Fulfillment flagged on, so that when a sentence is traced back to that intent, a request is propagated to the address where the JaCaMo agent is listening. It is relevant to remind here that only intents with the Fulfillment flagged on are propagated to the JaCaMo agent; all other intents are not propagated, which means are not in need of an agent to be handled (e.g., sentences not concerning the scene).


After the \textit{AddObject} intent has been generated by DialogFlow and propagated to the \vesna agent along with the \textit{who} and \textit{where} information, a request to Unity via http is sent by the agent.

Unity has an asynchronous listener waiting for requests and receives all the information sent by the agent as strings.
For example, if the user said in the chat \textit{``Put that object on the right''}, what arrives to Unity would only be \texttt{``that object, center, right''}.
After that, Unity converts the received string into a vector that describes the object position inside the scene. In case of global positioning, the resulting conversion is simple and is computed w.r.t. the size of the scene. However, for relative positioning, Unity has to look for the object used as a reference to compute the new position w.r.t the latter.

Once the position for the new object is found, a physical check is performed on the Unity side to be sure the selected place has not already been taken by another object (this is done by exploiting native Unity's colliders). If the position is already taken, an error message is sent back to the JaCaMo agent. Otherwise, the object is added and a ``done'' message is sent back to the JaCaMo agent containing the unique name of the object just added.

At this point, JaCaMo sends back a positive answer to DialogFlow, and a message is displayed to the user via the chat. Finally, the user may decide to stop or to start another iteration by adding a brand new object to the scene.

\begin{figure}[ht]
	\centering
	\begin{subfigure}{.4\textwidth}
		\centering
		\includegraphics[scale=.15]{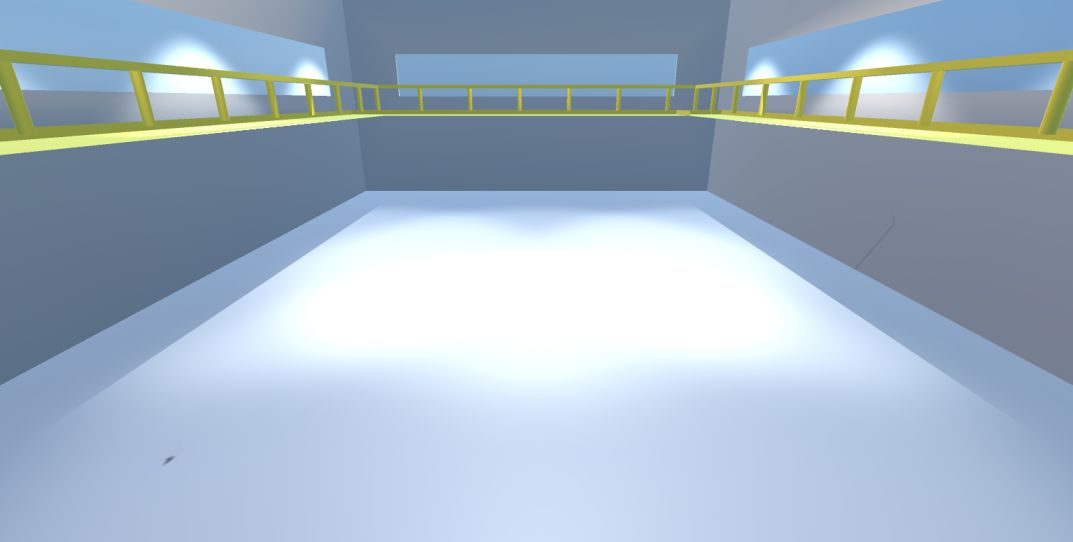}
		\subcaption{The factory}
		\label{factory}
	\end{subfigure}
	\begin{subfigure}{.4\textwidth}
		\centering
		\includegraphics[scale=.15]{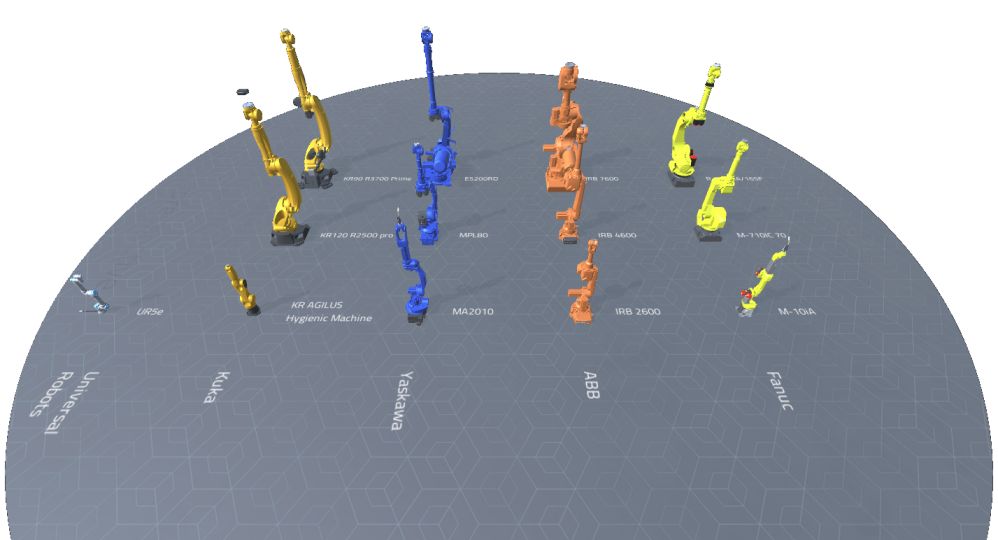}
		\subcaption{The robots}
		\label{robots}
	\end{subfigure}
	\caption{Model in Unity.}
\end{figure}

As a concrete example of use, let us consider the scenario introduced in Section \ref{introMot}. Let us suppose that \vesna's user is the owner of a small-medium factory where extended-reach welding robots (robots that can move their only arm in almost all directions, but cannot wander through the factory, see Figure \ref{robots}) must be positioned in an optimal way to carry out some automation work.
When \vesna is first run, an empty model of the factory in Unity is available to the user, ready to be modified (Figure \ref{factory}).
In this case, the user is interested in simulating in Unity the outcome of putting robots in different positions, to devise the optimal one. However, the user might not have the programming skills for working in Unity.
This is when \vesna comes into play to let the user position objects in the Unity scene by using natural language.
Let us suppose that the user wants to position a Yaskawa MA2010 robot\footnote{\url{https://www.robots.com/robots/motoman-ma2010}.} in the front-right global position. The user may type \textit{"Add a Yaskawa MA2010 in front on the right"} in the \vesna chat. 
\begin{figure}[ht]
	\centering
	\begin{subfigure}{.25\textwidth}
	\centering
		\includegraphics[scale=.3]{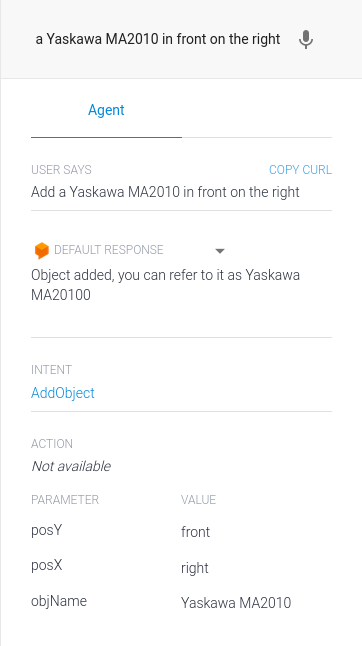}
		\subcaption{The Dialogflow chat}
		\label{dialogAdd}
	\end{subfigure}
	\begin{subfigure}{.6\textwidth}
	\centering
		\includegraphics[scale=.15]{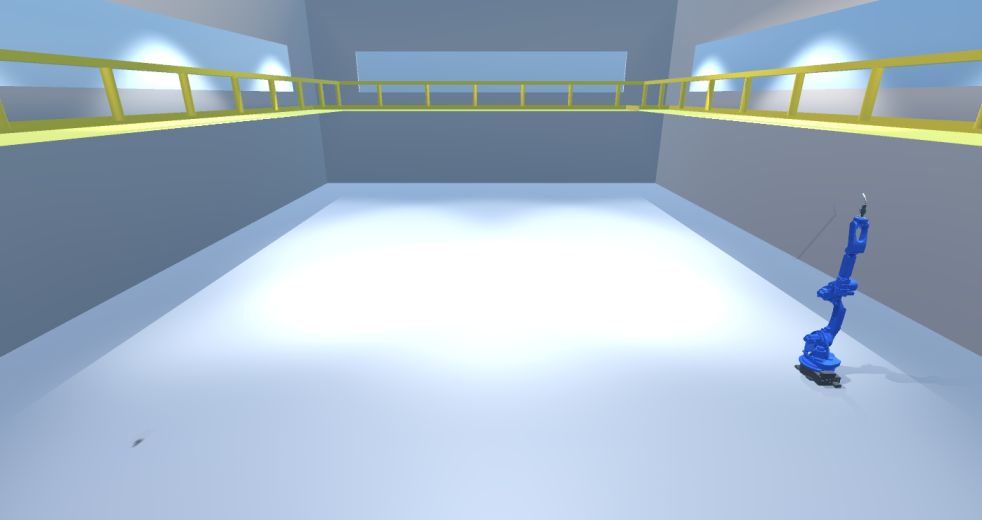}
		\subcaption{The graphical result}
		\label{factoryMA2010}
	\end{subfigure}
	\caption{Example of object addition with global coordinates.}
\end{figure}
The sentence in the chat (Figure~\ref{dialogAdd}) automatically creates the scene in Figure~\ref{factoryMA2010}.
The machinery behind the result displayed in Figure~\ref{factoryMA2010} is the following.
The Dialogflow agent brings the \textit{"Add a Yaskawa MA2010 in front on the right"} back to the intent \textit{AddObject}.
An entity \textbf{Object} that has parameters
\begin{enumerate}
	\item \textbf{Name}, the name of the object to be positioned, the robot in this case;
	\item \textbf{PosX}, in global positioning is the horizontal position, in relative positioning contains the relative position (\textit{left of}, \textit{right of}, \textit{behind}, \textit{in front of});
	\item \textbf{PosY}, in global positioning is the forward position, in relative positioning contains the name of the object the user referes to.
\end{enumerate}
The \vesna agent finds in the sentence the name \textit{"Yaskawa MA2010"}, the posX \textit{"right"} and the posY \textit{"front"} and sends a request to the link provided in the Fulfillment (so where the JaCaMo agents awaits). 

Dial4Jaca receives the information from Dialogflow and adds the following belief to the JaCaMo agent:
\begin{lstlisting}
request(
	"undefined",
	"5b485464-f275-42ab-853e-59514b115359-cf898478",
	"AddObject",
	[
		param("posX","right"),
		param("posY","front"),
		param("objName","Yaskawa MA2010")
	],
	...
\end{lstlisting}


The agent gets the object name and the position, then it calls the \texttt{addObject} function, implemented in a JaCaMo artifact.
This function sends an HTTP request to the 8081 port on the same machine and waits for an answer (we remind that on port 8081 Unity is listening for messages).
The HTTP request is assembled as follows
\[
	\texttt{http://localhost:8081/}\underbrace{\texttt{Yaskawa\%20MA2010}}_{\texttt{obj name}}\texttt{/}\underbrace{\texttt{right}}_{\texttt{posX}}\texttt{/}\underbrace{\texttt{front}}_{\texttt{posY}}
\]
On the Unity side, the script attached to the factory has a listener on that port and receives this piece of information.
It then converts the location from a couple of strings (in our example $(right, front)$) into a 3D vector of coordinates.
When it has a position, it checks that it is not already occupied by something else and adds it to the scene.

Let us now assume that the user wants to continue the interaction with \vesna. For instance, the user could ask to add a new robot to the left of the one inside the scene. Note that, since we now have already another robot inside the scene, we can add objects and position them by reference.
In this case, the user could write on the chat something like \textit{``Add a ABB IRB 2600 left of Yaskawa MA2010''}. The result we would obtain by such interaction is shown in Figure~\ref{factoryIRB2600}:
\begin{figure}[ht]
	\centering
	\begin{subfigure}{.25\textwidth}
	\centering
		\includegraphics[scale=.3]{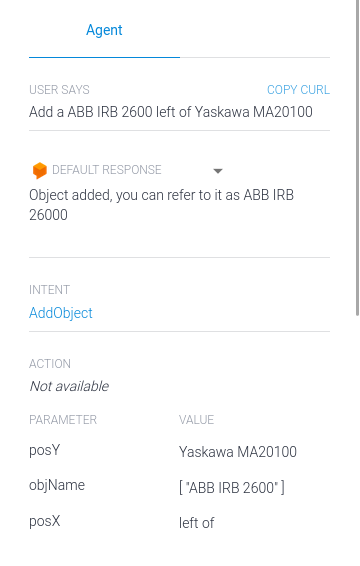}
		\subcaption{The DialogFlow chat}
		\label{dialogAddReference}
	\end{subfigure}
	\begin{subfigure}{.6\textwidth}
	\centering
		\includegraphics[scale=.15]{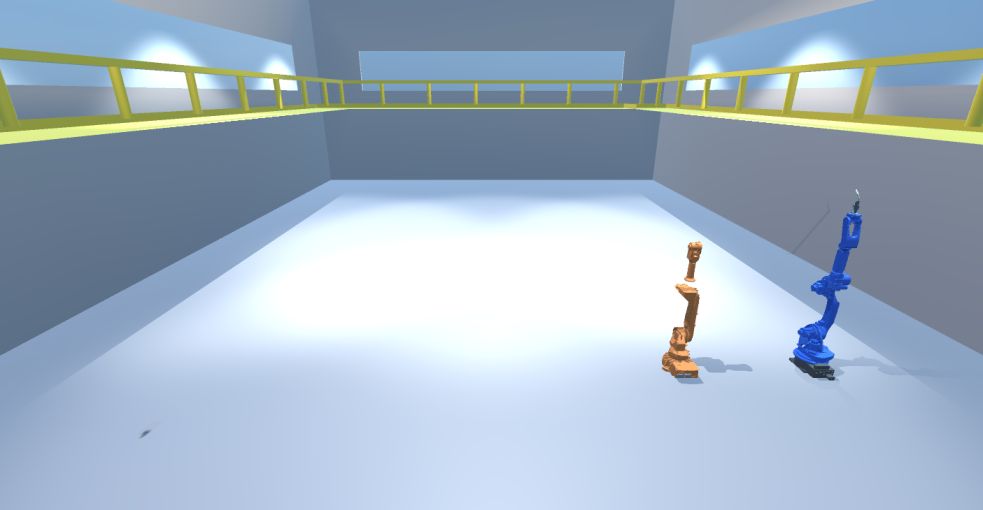}
		\subcaption{The graphical result}
		\label{factoryIRB2600}
	\end{subfigure}
	\caption{Example of object addition with relative coordinates.}
\end{figure}

\begin{figure}[ht]
    \centering
    \includegraphics[scale=.16]{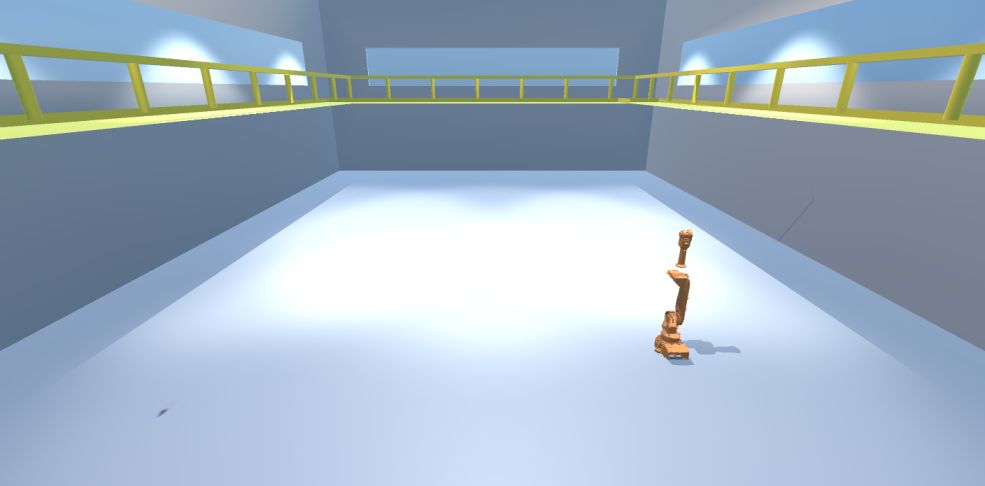}
    %
    \includegraphics[scale=.16]{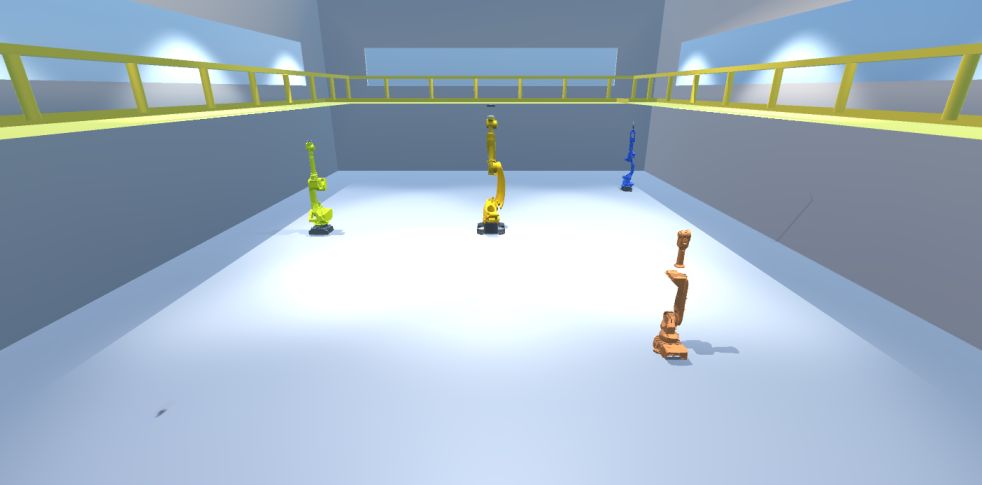}
    		\caption{Example of object removal (left) and result of further interactions (right).}
\label{remove-final}
\end{figure}

As last example, let us suppose that the user changes his/her mind and decides to remove the robot positioned at the beginning.
The user could type \textit{"Remove the Yaskawa MA20100"} and the result would be as shown in Figure~\ref{remove-final}, left, where the blue arm is no longer part of the scene.

Finally, after a few iterations, all the robots would be positioned in the factory as the user desires and the result would look like the scene shown in Figure~\ref{remove-final}, right (this is just a possible example of final result).


\section{Conclusions and Future Work}
\label{sec:conclusions-future-work}

%

We presented \vesna, a general-purpose framework to allow any users to populate simulations in Unity through natural language, heavily exploiting agent-based technologies that might help in reasoning on scenarios, verifying the feasibility of choices, providing explanations. We showed how starting from natural language sentences, \vesna can automatically handle the creation, addition, and removal of objects into a simulated scene in Unity. Specifically, this is obtained in \vesna through the combination of three different frameworks: DialogFlow, JaCaMo, and Unity. These frameworks support natural language analysis, the scene construction, and its graphic representation, respectively.
To the best of our knowledge, \vesna is the first framework which combines all these three aspects together.

For future developments, we are planning to explore additional uses for the objects in the simulated scene. In particular, instead of having static objects, we might have active objects as agents. This would help making the simulation more realistic, engaging, and amenable for a more precise risk analysis. In the factory automation scenario, employees could also be simulated as agents in the factory to experiment the movements and the interactions with objects and agents therein.

Being general-purpose, \vesna is not limited to the factory automation domain. Indeed, one of our original reasons for the development of \vesna was to create a tool for interactive, agent-based theatrical story-telling, along the lines of \cite{SPIERLING200231,10.1007/978-3-642-10291-2_44,rank12012creativity,10.1007/978-3-540-89454-4_13,berov2017character}. In that domain the presence of cognitive, goal-driven  characters in the Unity scene able to perform some actions on their own, provides a strong motivation for the integration of JaCaMo and Unity.

Finally, from a more practical perspective, we are also considering to explore additional frameworks to improve the \vesna usability and accessibility. For instance, other options are available on the sentence analysis side, such as Rasa\footnote{\url{https://rasa.com/}}. Rasa is a valid alternative to DialogFlow because open-source and can be run locally, differently from DialogFlow which instead requires to be executed in the cloud. At the current state, we opted for DialogFlow for its simplicity, and native support in Dial4JaCa. However, if used in scenarios where privacy might be an issue, the possibility to have all software running locally could be a necessary feature.

\nocite{*}
\bibliographystyle{eptcs}
\bibliography{bibliography}

\end{document}